\documentclass[prl, aps, amsfonts, amsmath, amssymb, twoside, twocolumn, superscriptaddress, sort&compress, reprint, 10pt]{revtex4-2}

\usepackage[latin1]{inputenc}
\usepackage[UKenglish]{babel}

\usepackage{nicefrac, graphicx, textcomp, mathtools, siunitx}

\newcommand{\ops}[1]{\hat{#1}}
\newcommand{\company}[1]{\textsf{#1}}
\newcommand{\Qf}{\ensuremath{\mathcal Q}}
\newcommand{\dint}[2]{\;\ifnum1=0#1%
\mathrm d #2\else\mathrm d^#1 #2\fi}

\DeclarePairedDelimiter\robra{\lparen}{\rparen}
\DeclarePairedDelimiter\abs{\lvert}{\rvert}

\begin{document}

\title{Commercial scanning nitrogen vacancy magnetometer in a closed-cycle cryostat}

\author{Clemens Sch\"afermeier}
\email{clemens.schaefermeier@attocube.com}
\affiliation{attocube systems AG, Eglfinger Weg 2, 85540 Haar, Germany}

\author{Christopher Kelvin von Grundherr}
\affiliation{attocube systems AG, Eglfinger Weg 2, 85540 Haar, Germany}

\author{Patrick Ebermann}
\affiliation{attocube systems AG, Eglfinger Weg 2, 85540 Haar, Germany}

\author{Dominik Irber}
\affiliation{attocube systems AG, Eglfinger Weg 2, 85540 Haar, Germany}

\author{Khaled Karra\"i}
\affiliation{attocube systems AG, Eglfinger Weg 2, 85540 Haar, Germany}

\author{Andrea Morales}
\affiliation{QZabre AG, 8050 Zurich, Switzerland}

\author{Jan Rhensius}
\affiliation{QZabre AG, 8050 Zurich, Switzerland}

\author{Gabriel Puebla-Hellmann}
\affiliation{QZabre AG, 8050 Zurich, Switzerland}

\begin{abstract}

The ability to measure magnetic fields on the nanometre scale at cryogenic temperatures is key to understand magnetism on the quantum level and to develop materials for new storage devices or quantum computers.
Nitrogen vacancy (NV) centres in diamond have proven to be a robust means to harness quantum sensing for such applications.
We have developed an instrument to measure the magnetic stray field of a sample with nanometre resolution from \SI{2}{\kelvin} to \SI{300}{\kelvin} and that accepts samples without additional preparation, especially the need to prepare a microwave line on the sample.
The instrument features a software-interface for controlling and synchronising all included optical, mechanical and electronic devices and which analyses the acquired information in real time.

We present the key features and measurement results achieved with atomic force microscopy (AFM) tips hosting an NV centre and a fully remote controllable microscope platform.
The magnetometer is commercially available, and a first demonstrator has been installed in a research facility.
We show \si{\micro\tesla\per\sqrt\hertz} sensitivity, low noise AFM tip control and optically detected resonance scans in a closed-cycle cryostat.

\end{abstract}

\maketitle

\section{Introduction}

Negatively charged nitrogen vacancy (NV) centres in diamond are atomically sized defects \cite{Doherty2013}.

The robust quantum properties of NV centres have been a subject of research for decades and make it appealing for quantum sensing applications \cite{Schirhagl2014, Marchiori2022, Budakian2024, Christensen2024}.
A key application is the quantitative detection of magnetic fields by means of optically detected magnetic resonance (ODMR).

To use NV centres for nanometre-scale sensing of magnetic fields, NV centres in nano diamonds have been attached to atomic force microscopy (AFM) tips \cite{Balasubramanian2008, Rondin2012, Tetienne2012, Rondin2013, Tetienne2013, Tetienne2014, Tetienne2015} or by using a diamond AFM tip hosting a single NV centre \cite{Maletinsky2012, Appel2015, Gross2017}.
Today, diamond AFM tips with a single NV centre, along complete microscope devices, are commercially available and are used in sensing applications.
The result of our work is the first commercially magnetometer with a validated magnetic scan at cryogenic temperatures.

As the NV centre is capable of sensing magnetic fields also at cryogenic temperatures, and studying magnetic fields at such temperatures is particularly important for novel materials of quantum devices, 2D materials, and fundamental aspects of magnetism, we have developed a cryogenic NV magnetometer.
In contrast to demonstrations of scanning NV magnetometer at cryogenic temperatures \cite{Thiel2016, Pellicione2016, Vool2021, Sun2021, Jenkins2022, Scheidegger2022, Tan2024}, our magnetometer does not require a sample prepared with a microwave antenna or strip line.
As the preparation with a microwave emitter typically restricts the addressable sample range, the removal of this requirement improves the ease of use significantly.
The cryostat is a closed-cycle system, which implies continues operation at temperatures down to \SI{1.7}{\kelvin} and decreases running costs by eliminating the need for refilling helium.

In this whitepaper, we present our key results and steps towards reliable low-temperature magnetometry.

\section{Results}

\begin{figure}
  \includegraphics[width = \columnwidth, clip, trim = 0 0 25 65]{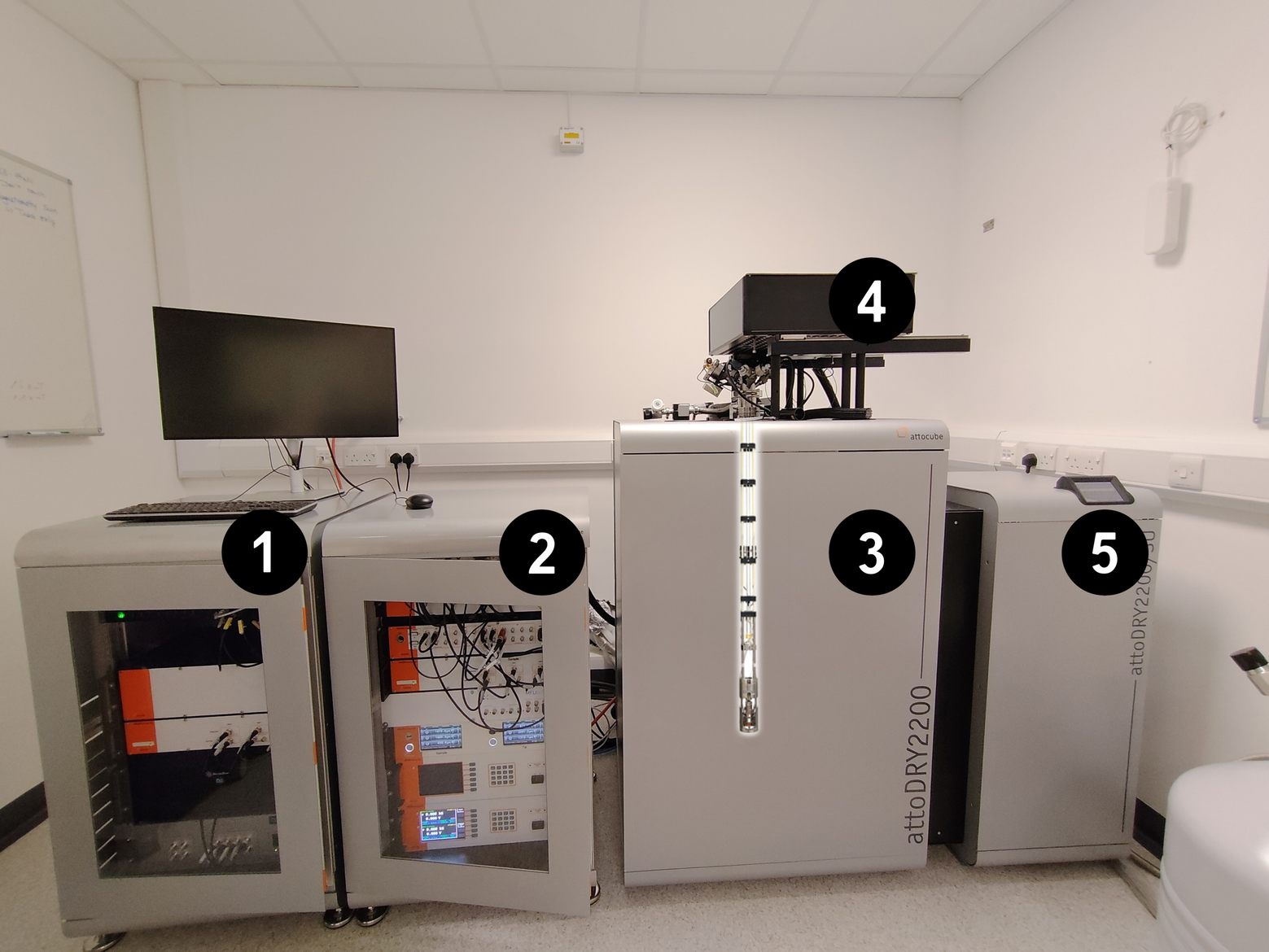}
  \caption{1: rack with computer, microwave generator, data acquisition, laser and photon counter.
  2: rack with controllers for AFM, positioners, and vector magnet.
  3: closed cycle cryostat with vector magnet and an illustration of the microscope stick that contains the sample, objective and the AFM tip.
  4: optics module for widefield imaging and confocal excitation and detection; motors to drive and control optomechanics included.
  5: cryostat support unit. Not shown: helium compressor unit.}
  \label{fig:system_photo}
\end{figure}

\begin{figure}
  \includegraphics[width = \columnwidth, clip, trim = 0 10 0 20]{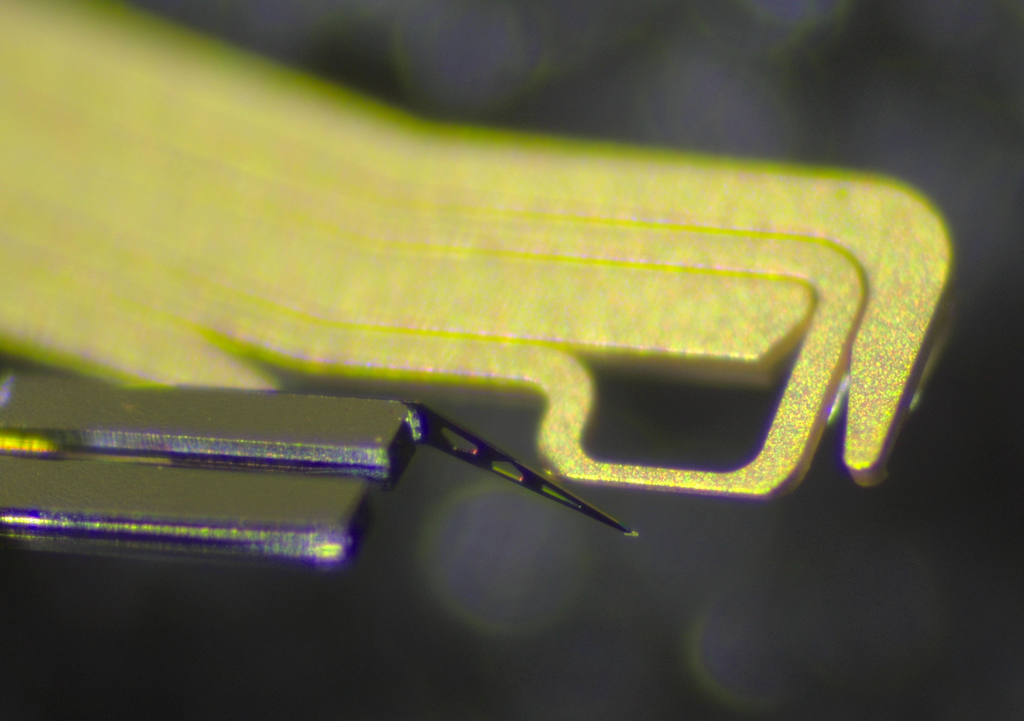}
  \caption{A photo of the tuning fork with a diamond tip and an integrated microwave line, fabricated by \company{QZabre}.}
  \label{fig:TFImages}
\end{figure}

The system is shown in figure \ref{fig:system_photo}.
All electronics and optics used for the measurement are mounted in two 19" racks.
The attoDRY2200 cryostat features a closed-cycle helium circulation and a guaranteed base temperature of sub \SI{1.8}{\kelvin}.
Due to a pneumatic damping system, the attoDRY2200 is a unique ultra-low vibration measurement platform for cryogenic scanning probe experiments.
The damping system efficiently decouples mechanical vibrations -- created internally by the pulse-tube coldhead and externally by other equipment and work -- from the sample space.
A superconducting vector magnet is available in different configurations and produces fields up to \SI{9}{\tesla} in one direction.
The cryostat hosts an optics module providing confocal and widefield imaging.
Widefield imaging is especially useful to locate the diamond tip and sample areas to study.
The field of view of the widefield imaging is \SI{55}{\micro\meter}, constrained by the aperture of the vacuum window.
To counteract drift due to thermal changes in the surrounding, two mirrors are remote controlled.
An alignment mode can be remotely activated to check for misalignment and optimises the detection efficiency when no widefield imaging is required.
The optics module is mounted on a roller bearing rail system which provides positional repeatability of \SI{20}{\micro\meter} on the sample space.
This means, that the user will not have to re-align the optics when the microscope stick is removed and inserted back into the cryostat, which has to be done for a sample exchange.
The sample and AFM tip holder can be translated individually by means of slip stick positioners with a range of \SI{3}{\milli\meter}, and by means of piezo scanners over a range of \SI{30}{\micro\meter} and \SI{15}{\micro\meter} at room and base temperature, respectively.
An apochromatic, low temperature and high vacuum compatible, objective is used to excite and detect the NV centre fluorescence.
The full-width-half-maximum of the point spread function of the \company{attocube} LT-APO objective measures \SI{468 \pm 2}{\nano\meter} and \SI{474 \pm 2}{\nano\meter} on the major and minor axis, respectively.
To perform ODMR, microwave excitation is required.
\company{QZabre} offers probes with an integrated microwave line in close vicinity to the NV centre, improving efficiency and eliminating the potentially harmful tip-line alignment typically necessary.
The AFM tip and the microwave line is shown in figure \ref{fig:TFImages}.
The integrated MW line can generate Rabi rates well exceeding \SI{10}{\mega\hertz} at the NV centre.

To perform a magnetometry scan, two steps have to be taken: AFM calibration and control, and characterisation of the NV centre.

To calibrate the AFM, we measure the tuning fork resonance caused only by Brownian force.
This step is important to relate the detected voltage to a displacement and force of the diamond tip.
Too much force can be harmful for delicate samples.

The AFM is operated by driving the shear mode of the tuning fork, which is at \SI{32.3 \pm .2}{\kilo\hertz}.
The charge generated due to the shear motion of the tuning fork is amplified by means of a low noise charge amplifier by \company{FEMTO} with \SI{1e13}{\volt\per\coulomb} gain.
\begin{figure}
  \includegraphics[clip, trim = 10 10 10 0]{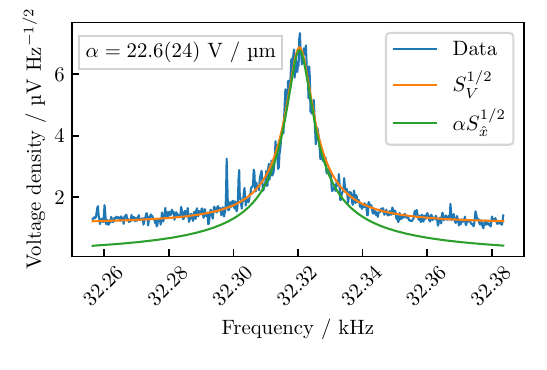}
  \caption{Tuning fork spectrum caused by Brownian force at room temperature.
  The measurement serves to calibrate the displacement and force of the tip.}
  \label{fig:BrownianForceCal}
\end{figure}
A lock-in amplifier by \company{Zurich Instruments} is used to analyse the charge amplifier's output.
Figure \ref{fig:BrownianForceCal} shows a typical tuning fork spectrum caused by Brownian force.
The measured power spectral density is fitted to \cite{Hauer2013}
\begin{equation}
  S_V(f) = \alpha^2 S_{\ops{x}}(f) + S_\text{eN},
  \label{eq:fitfun}
\end{equation}
where $\alpha$ is the transduction in units of \si{\volt\per\micro\meter}, $S_\text{eN}$ is the electronic noise floor of the setup and
\begin{equation}
  S_{\ops{x}}(f) = \frac{k_B T f_\text{res}}{2 \pi^3 m_\text{eff} \Qf ((f^2 - f_\text{res}^2)^2 + (f f_\text{res} / \Qf)^2)},
\end{equation}
where $k_B$ is Boltzmann's constant, $T$ the temperature, the natural frequency $f$, $f_\text{res}$ as the resonance frequency, $\Qf$ the Q factor of the tuning fork, and $m_\text{eff}$ the effective mass of the shear mode.
To calculate the effective mass, we performed the integral $m_\text{eff} =\int \dint{1}{V} \rho(\vec x) \abs{\vec{r}(\vec x)}$ over a numerical model of the tuning fork.
In the integral, $\vec r$ is the normalised displacement vector and $\rho$ the density.
If $m_\text{eff}$ determined numerically, $T$ is known, $S_\text{eN}$ is measured, then equation \eqref{eq:fitfun} contains only three free parameters: $\Qf$, $f_\text{res}$ and $\alpha$.
To validate our approach, we compared the outlined procedure to an optical measurement.
First, a Fabry--P\'erot interferometer was built to measure the displacement of the tip of a tuning fork prong.
Then, the tuning fork was driven at its shear mode resonance frequency by means of a ``dither'' piezo attached under the tuning fork.
The optical measured yielded \SI{2.47 \pm 0.08}{\nano\meter}.
Finally, we used the Brownian force calibration method as described above and converted the peak amplitude of the excited tuning fork with the transduction factor $\alpha$.
The result is \SI{2.32 \pm 0.24}{\nano\meter}.
While the uncertainty of the calibration method is larger than the optical measurements, the Brownian force calibration is faster and work ``in vivo''.
As the spring constant of the tuning fork and cantilever of the diamond tip is $\approx \SI{10}{\newton\per\meter}$, the force generated here is $\approx$ \SI{23}{\nano\newton}.
The force generated by the tuning fork can be tuned directly by the voltage that drives the dither piezo.
In a typical measurement the tuning fork displacement amplitude is $\approx$ \SI{1}{\nano\meter} which results in a force of $\approx$ \SI{10}{\nano\newton}.

After the tuning fork is calibrated, the tip-sample interaction can be probed for adjusting loop parameters of the AFM controller.
\begin{figure}
  \includegraphics[clip, trim = 10 10 10 0]{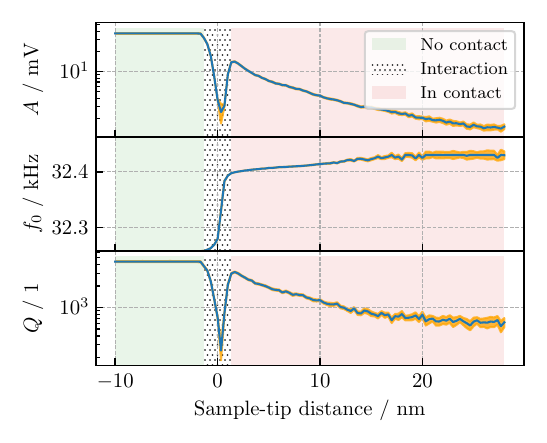}
  \caption{Approach curve of the tuning fork on a SiN sample (\company{Anfatec} UMG01).
  At each sample-tip distance, a tuning fork resonance sweep was performed and the amplitude $A$, resonance frequency $f_0$ and quality factor $Q$ extracted.
  The orange area above and below each line visualises the standard deviation of the estimate.
  A positive distance means that tip and sample are pressing against each other.}
  \label{fig:approach_curve}
\end{figure}
This is achieved with an approach curve, where the tip-sample distance is minimised and at each distance, a tuning fork spectrum is taken \cite{Grober2000, Karrai2000}.
The \company{Zurich Instruments} MFLI integrated in our device is programmed for control and measure this process.
The data extracted from each tuning fork spectrum is shown in figure \ref{fig:approach_curve}.
The plot shows the change in peak amplitude, resonance frequency and Q factor over the tip-sample distance.
Knowing the frequency shift over sample distance is the key to set up a phase-lock-loop AFM control without guessing \cite{Gildemeister2007}.

We chose to plot a negative distance where tip and sample are not in contact.
The distance is zeroed where the frequency shift is strongest, as it is the range where the tip and the sample are interacting, but are not in contact.
Beyond a zero distance, the sample can be ``pressed'' further into the tip (or vice versa), which is identified in the plot as a positive distance.
In the plot, the interacting distance measures less \SI{3}{\nano\meter}.
First of all, it proves that the cryostat is capable of delivering low-vibration performance, second, it means that in this case, fast feedback operation is necessary for stable AFM measurements.

Magnetometry measurements can be taken in all three regimes: above the sample, in interaction and in physical contact.
To have a high resolution, a close distance is better than further away.
On the other other, strong off-axis fields can lead to quenching of the ODMR, which can be mitigated by hovering above the sample.

With the AFM part characterised and set up, we showcase the optical properties of the NV centre in the diamond tip.
Our motivation is to optimise the magnetic field sensitivity as it speeds up the measurement time and improves the magnetic resolution.
To understand the importance of the optics, we review the magnetic field sensitivity.
The DC magnetic field sensitivity in units of \si{\tesla\per\sqrt\hertz} is approximated by \cite{Rondin2014Review}
\begin{equation}
  \sigma \approx \frac{h}{g \mu_B} \frac{\Delta \nu}{C \sqrt r},
\end{equation}
where $h$ is Planck's constant, $g$ the gyromagnetic ratio, $\mu_B$ the Bohr magneton.
$C$ is the ODMR contrast and intrinsic to the NV centre and its surrounding, thus hardly influenced by the setup.
$\Delta \nu$ is the ODMR linewidth, limited by inhomogeneous dephasing time $T_2^*$.
This fundamental limited may be reached by ``pulsed'' sensing protocols, where preparation, magnetic field measurement, and readout is time-separated.
There are various protocols implemented in our device to do that.
The last parameter is the countrate $\sqrt r$ which is limited by the detection efficiency of the fluorescence.
As there are various pulsed protocols available, we focus here on the detection efficiency to provide a benchmarking for the user.
Given a single NV centre is present in the tip, a fact that is established during production, the best benchmark is a saturation measurement.
\begin{figure}
  \includegraphics[clip, trim = 10 10 10 0]{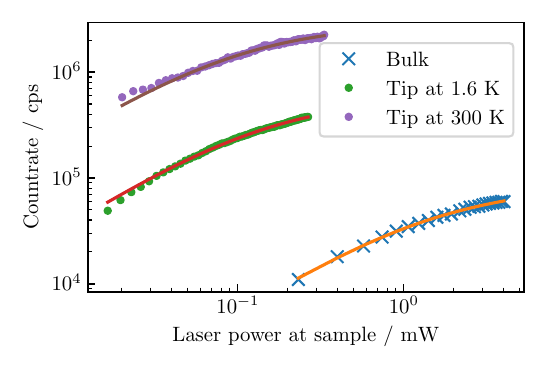}
  \caption{Saturation measurements of a single NV centre at \SI{300}{\kelvin} in bulk diamond, compared to an NV centre in a \company{QZabre} tip at \SI{300}{\kelvin} and \SI{1.6}{\kelvin}.
  The power at the sample was determined experimentally prior to the saturation measurement.
  }
  \label{fig:sat_plot}
\end{figure}
Figure \ref{fig:sat_plot} shows the optical performance of the instrument at \SI{1.6}{\kelvin} and \SI{300}{\kelvin} by means of a saturation measurement.
In addition, a saturation curve of a single NV centre in bulk diamond, \SI{400}{\nano\meter} below the diamond surface, is shown as reference.
All three saturation curves were measured with our system.
First, the optical power was measured directly at the focal plane of the objective.
Next, the NV centre was located and its position optimised.
After taking the fluorescence data, we performed a fit to $r_\infty p / \robra{p + p_\text{sat}} + a p + b$, where $r_\infty$ is the countrate in the steady state under continuous excitation, $p_\text{sat}$ the saturation intensity and $a$ and $b$ is the linear approximation of background counts and dark counts, respectively, which were both determined experimentally.
\begin{table}%
  \begin{tabular}{l | c c}
    NV centre in\dots         & $r_\infty$ / \nicefrac{$10^3$}{s} \qquad & $p_\text{sat}$ / \si{\micro\watt} \\ \hline
    bulk at \SI{300}{\kelvin} & \num{79 \pm 1}                           & \num{1444 \pm 42} \\
    tip at \SI{300}{\kelvin}  & \num{2591 \pm 146}                       & \num{100 \pm 10} \\
    tip at \SI{1.6}{\kelvin}  & \num{579 \pm 56}                         & \num{146 \pm 18} \\
  \end{tabular}
  \caption{Saturation parameters of the measurements shown in figure \ref{fig:sat_plot}.}
  \label{tab:sat_rates}
\end{table}
Table \ref{tab:sat_rates} summarises the parameters $r_\infty$ and $p_\text{sat}$ for the three saturation measurements.
Benchmarking the room temperature rates in bulk and the AFM tip against literature shows a very good performance of the collection efficiency \cite{Riedel2014}.

As a last step, we quantify the magnetic field sensitivity.
\begin{figure}
  \includegraphics[clip, trim = 10 10 10 0]{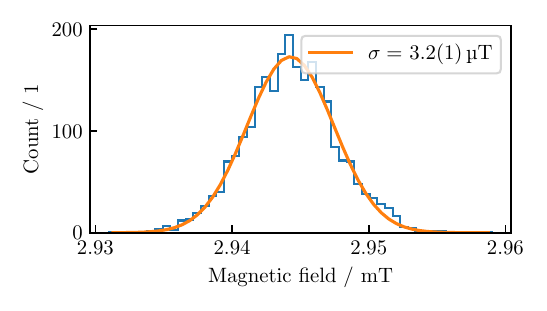}
  \caption{Magnetic field sensitivity at \SI{2.8}{\kelvin}.
  The bin size is \SI{0.56}{\micro\tesla}; \num{2500} measurements with a duration of \SI{1}{\second} per measurement were performed.
  The sample was \SI{1}{\milli\meter} below the NV centre, the vector magnet was set up to provide the bias field.}
  \label{fig:mag_sens}
\end{figure}
For showcasing, we performed an ODMR, far away from a sample, over \SI{1}{\second} and evaluated the distribution of the results.
Figure \ref{fig:mag_sens} shows that we achieve a DC magnetic field sensitivity of \SI{3.2 \pm .1}{\micro\tesla\per\sqrt\hertz}.
Due to local heating of the microwave excitation, the temperature increased to \SI{2.8}{\kelvin} during the measurement.

Finally, we present a magnetic field measurement combining all ingredients.
As a sample we used a Ir/Fe/Co/Pt multilayer sample.
Due to its strong off-axis magnetic field, the ODMR contrast quenches in contact with the sample.
That is why we measured with a constant height above the sample.
\begin{figure}
  \includegraphics[clip, trim = 10 10 10 0]{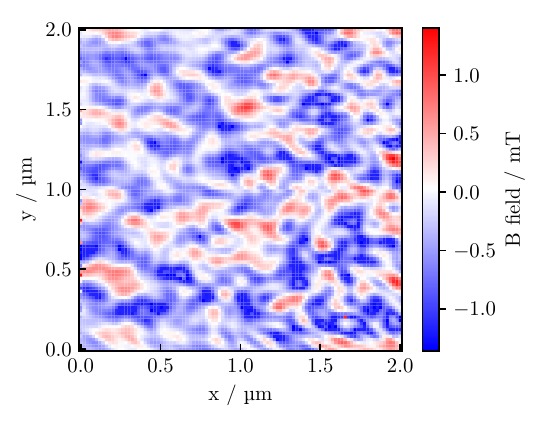}
  \caption{Magnetic field measured on a Ir/Fe/Co/Pt multilayer sample by means of CW ODMR.
  During the scan the sample temperature was \SI{2.9}{\kelvin}.
  The tip was set to a constant offset of \SI{50}{\nano\meter} above the sample.
  The pixel-pixel distance is \SI{20}{\nano\meter}.
  No post-processing was performed.}
  \label{fig:odmr_scan}
\end{figure}
The height was set by first establishing AFM contact to the sample, then retracting the sample by \SI{50}{\nano\meter}.
Figure \ref{fig:odmr_scan} shows the non-post-processed magnetic field distribution generated by the sample acquired at \SI{2.9}{\kelvin}.
The CW ODMR scan with a \SI{20}{\nano\meter} step size took \SI{8}{\hour} and required no manual correction.

An important quality of the device is the lateral spatial resolution of the magnetic field.
The resolution is approximately limited by the distance between the NV centre and the sample.
It is important to note that this limit also includes the distance between the NV centre and the facet of the diamond tip; which is, thanks to latest advances in the fabrication of the diamond tips, approximately \SI{10}{\nano\meter}.
\begin{figure}
  \centering
  \includegraphics[clip, trim = 10 10 10 0]{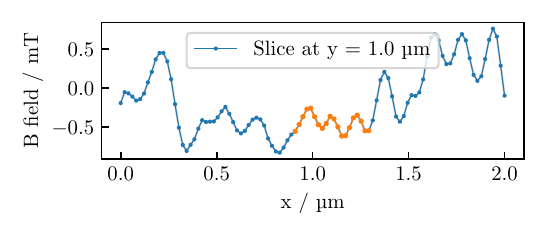}\\
  \includegraphics[clip, trim = 10 10 10 0]{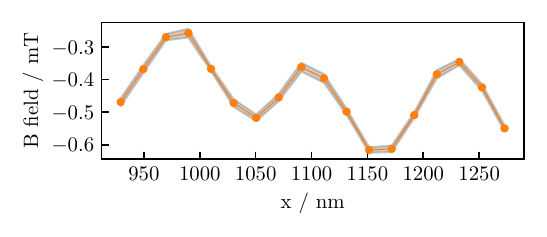}
  \caption{Top: Cross section of magnetic field of figure \ref{fig:odmr_scan}.
  Bottom: Zoom into cross section.
  The grey band visualises the standard deviation of each data point.}
  \label{fig:odmr_crosssecBfield}
\end{figure}
To assess the lateral resolution which is a convolution of the probe and the domain wall width, figure \ref{fig:odmr_crosssecBfield} plots a cross section of the magnetic field map.
Fitting a Gaussian distribution to the peaks shown in the cross section, one find a resolution in terms of a standard deviation of \SI{33 \pm 1}{\nano\meter}, equivalent to a full width half maximum of \SI{77}{\nano\meter} and is in good agreement with the resolution limited by the total distance of approximately \SI{65}{\nano\meter} between NV centre and sample surface.

\paragraph{Conclusion}

In conclusion, this is the first paper that presents the development and successful operation of a commercial scanning nitrogen vacancy magnetometer in a closed-cycle cryostat.
The instrument demonstrates high sensitivity and low-noise performance, and is capable of measuring magnetic fields at the nanometre scale from \SI{2}{\kelvin} to \SI{300}{\kelvin} without the need for additional sample preparation.
The integration of advanced optical, mechanical, and electronic components, along with the user-friendly software interface, ensures precise, non-invasive, and efficient quantitative magnetic field measurements.
Our instrument changes the paradigm for novel research and applications in quantum sensing and materials science, by providing tangible improvements in precision and time-to-result for this powerful magnetic imaging technique.

\paragraph{Acknowledgement}

\company{attocube} acknowledges support from the EU flagship program ASTERIQS.
We are grateful for Saddem Chouaieb's hands-on support during the experimental phase.
The sample for magnetic imaging was provided by Anjan Soumyanarayanan from the Institute of Materials Research and Engineering in Singapore.

\end{document}